\documentclass{article}
\title{Quantum mechanics without interpretation}
\author{O. Yaremchuk\\P.N. Lebedev Physical Institute,
RAS\\
email: yarem@online.ru}
\date{\today}
\begin{document}
\maketitle
\begin{abstract}

It is shown that the independence of the continuum hypothesis
points to the unique definite status of the set of intermediate
cardinality: the intermediate set exists only as a subset of
continuum. This latent status is a consequence of duality of the
members of the set. Due to the structural inhomogeneity of the
intermediate set, its complete description falls into several
``sections'' (theories) with their special main laws, dimensions,
and directions, i.e., the complete description of the one-dimensional
intermediate set is multidimensional. Quantum mechanics is one
of these theories.

\end{abstract}

The concept of discrete space is always regarded as a unique
alternative to continuous space. Nevertheless, there is one
more possibility originated from the continuum problem: the
set of intermediate cardinality represents the golden mean
between the opposing concepts.

But although the continuum problem has been solved \cite{Cohen},
the status of the set of intermediate cardinality is still unclear.
The commonly held view is that, according to the independence of
the continuum hypothesis (CH), we can neither prove nor refute
existence of the intermediate set and, therefore, CH or its negation
must be taken as an additional axiom. 

However, it may be stated that there exists a unique definite status
of the intermediate set which is consistent with the independence of the
continuum hypothesis.
Since, by definition, the set of intermediate cardinality $I$ should be a
subset of continuum $R$, the independence of CH may be understood as
impossibility, in principle, to separate the subset of intermediate
cardinality from continuum. 
In other words, the independence of CH means that for any real number
$x\in R$ the sentence $x\in I\subset R$ is undecidable.
Reasons of this inseparability, if any, should be investigated.

This latent status is the only definite status of the
set of intermediate cardinality that is consistent with the generally
accepted solution of the continuum problem. However, in order to state
that the status follows from the independence of CH, it is necessary to
know that standard Zermelo-Fraenkel set theory (ZF) gives correct
description of the notion of set, i.e., we need set-theoretic analog of
Church's thesis in order to be sure that we have reliable solution of the
continuum problem independent of the concrete formalization of the concept
of set. This thesis ensures that any changes in the collection of
self-evident statements will not alter the present solution of the
problem and therefore the Zermelo-Fraenkel collection of axioms is, in
this sense, complete. This understanding of completeness implies that the
independence (and the existence) of the statement is not an imperfection or
a peculiarity of the formalization but a sign of some property of the set
in question.

Note that, if we postulate existence of the intermediate set (in other
words, if we take the negation of CH as an axiom), the result will be the
same: since any way of obtaining the set is forbidden by the independence
of CH, we have to reconcile with the same hidden intermediate subset in
continuum which we can get without any additional assumption.

\medskip
In order to separate some finite set we do not need any fixed rule
(algorithm) if we are interested only in the number of members.
We may select members arbitrarily until the collection has requisite
cardinality. 
In the case of an infinite subset, the separation procedure must be based on
some appropriate rule. From the undecidability of CH it follows that we
do not have a rule for separation of the intermediate subset.

According to the separation axiom scheme, for any set and for any
property expressed by some formula there exists a subset of the set,
which contains only members of the set having the property. Hence, if
the members of the intermediate subset have any discriminating property
relative to the other members of continuum, then we can separate the
subset. From the independence of CH it follows that we cannot express
any property of the members of the intermediate set, i.e., any property
we can formulate implies separation of either countable or continuous
subset of continuum. This does not mean that members of the intermediate
set have no properties and therefore this subset is empty.

If we have a set consisting of members of two kinds, it is possible that
there are exist members combining properties of both kinds. This trivial
possibility makes substantial difficulties when the properties are mutually
exclusive.
In this case, the properties may not become apparent simultaneously.
However, they may manifest itself under different tests: such a dual
member ``looks'' exactly like a member of one of these kinds depending on
testing procedure. This complicates the problem: we must introduce two
alternative kinds of testing procedure and ensure comprehensive test of
each member. But set theory does not provide for the mutually exclusive
properties. Since one-stage test is not sufficient for detection of the
dual members, we get the independent statement. 

As an illustration, consider a brick road which consists
of black bricks and white bricks. If we know (or suspect)
that among them there are some bricks which have white
top sides and black bottom sides (or vice versa, i.e.,
these bricks combine the sides ``from'' the white brick
and ``from'' the black brick), we, nevertheless, cannot
find them. Based only on the top view, the problem of separation
(and even existence) of the black-and-white bricks of this kind
is undecidable. Each brick can be black-and-white with equal
probability. However, if we have the top view and the bottom
view, we can find these bricks: each of them looks like a white
brick in the one view and like a black brick in the other view
(``black-white duality'').

What is the basic difference between the members of the countable set and
continuum?
Here we touch upon the problem of relationship between cardinality and
nature of a set. George Cantor believed that cardinality of a set is
independent of nature of its members. In accord with this premise,
we can ignore the properties of the members of all sets and  the separation
axiom scheme. But, in this way, we can miss an opportunity to become
aware of one more important problem. The absence of correlation
between cardinality and structure of a set is neither axiom nor theorem.
This is obvious only for finite sets. For any infinite set, it is quite
clear that, ``to be a member of the set'' means ``to have some special
property''. Of course, we mean typical set-generating members. An infinite
set is always coupled with its natural structure which is, actually, the
optimal arrangement of its members. The members should be suitable for
embedding in the natural structure of the set.

Let us explicitly formulate the following second thesis: structure of an
infinite set is determined by its cardinality directly related to the
properties of its members.

Cardinality of a finite set may be characterized by most complex
structure which can be made of this number of members. On the contrary,
the distinctive structure of infinite cardinality is the simplest structure
(the most compact arrangement) of the corresponding set. For example,
the countable set can fit into the natural structure of continuum (the set of
the rational numbers) and there remain vacancies (the irrational numbers),
whereas continuum cannot be arranged in the natural structure of the
countable set where for any real number there is the only successor:
the discrete structure of the set of natural numbers cannot accommodate all
real numbers. It is clear that the most symmetrical arrangement of members
should be regarded as the simplest structure of a set.
 
The relationship between cardinality and structure of a set can be the
cause of some seemingly baseless phenomena: the minimization principles
(e.g. the principle of least action), spontaneous symmetry breaking,
phase transitions, etc. 

From the operational viewpoint, a natural number and a real number
are outputs of different procedures: counting and measurement. Any real
number is length of some interval or segment, i.e., the key property,
making the difference, is plausibly length considered as a qualitative
(physical) property ``to be extended''. Roughly speaking, the countable
set basically consist of the zero-length (point-like) members, while the
generic members of continuum have non-zero length. Length, as a
property, means applicability, as an ``assembly unit'', to the structure of
continuum. Corresponding dual member ``looks'' either like a point or like
an extended object depending on the testing arrangement.

In set theory, uncountable sets appear only due to the power set axiom,
i.e., the magic word ``$2^N$'' should produce an extended object from
nothing (the countable set is made of the empty set by axiom of pairs and
axiom of infinity).

Thus the only apparent reason for the inseparability of the intermediate
set is duality of its members (except for the members of the countable
subset of the intermediate set). But we get a problem of formulation of the
corresponding alternative properties and the testing procedures
in terms of set theory.

We confine discussion of this duality to the above informal remarks
(we shall not use it in further consideration). But it is worthwhile to
note that this duality is similar to wave-particle duality: in quantum
mechanics, we have continuous space of classical mechanics (a macroscopic
measuring apparatus described by classical rules) and, inside this
continuum, some dual objects (microscopic particles) combining alternative
properties of a wave (continuum) and a point-like particle (the countable
set). Without this similarity the above observations may seem too bizarre
and vague to be taken into account. But the direct suggestion of reality may
not be ignored.

A set is a real object and may have {\it a priori} unpredictable properties,
particularly, if it consist of very small (microscopic) or very large
(cosmological) members. This properties should be related, naturally,
to independent statements which indicates that we deal with more complete
reality than is fixed in the formal system based on the concepts originated
in our usual scale.

Recall that non-Euclidean geometry and general relativity have emerged from
the independence of the fifth Euclid's postulate. Analogously, the
independence of CH converts existence of the intermediate set into a
physical problem. The structure of space (space-time) is not only geometry.
Moreover, set-theoretic structure determines validity of geometry: geometry
is a set of properties of continuous space.

\medskip
Consider the maps of the intermediate set $I$ to
the sets of real numbers $R$ and natural numbers $N$.

Let the map $I\to N$ decompose $I$ into
the countable set of mutually disjoint
infinite subsets: $\cup I_n=I$ ($n\in N$).
Let $I_n$ be called a unit set. All members of $I_n$
have the same countable coordinate $n$.

Consider the map $I\to R$.
By definition, continuum $R$ contains a subset $M$ equivalent to $I$,
i.e., there exists a bijection
\begin{equation}
f:I\to M\subset R.
\end{equation}
This bijection reduces to separation of the intermediate
subset $M$ from continuum. Since any separation rule
is a proof of existence of the intermediate
set and, therefore, contradicts the independence
of the continuum hypothesis, we, in principle, do not
have a rule for assigning a real number to an
arbitrary point $s$ of the intermediate set. Hence, any
bijection can take the point only to a random real number.
If we do not have preferable real numbers,
then the mapping is equiprobable, i.e., the point can
be found at any real number with equal probability.
This already conforms to the quantum free particle.
In the general case, we have the probability $P(r)dr$ of
finding the point $s\in I$ about $r$. Note that we have used
only the independence of CH and the thesis about validity of ZF.

Thus the point of the intermediate set has two coordinates:
a definite natural number and a random real number:
\begin{equation}\label{s}
s:(n,r_{random}).
\end{equation}
Only the natural number coordinate gives reliable
information about relative positions of the points
of the set and, consequently, about size of an interval. 
But the points of a unit set are indistinguishable. It is
clear that the probability $P(r)$ depends on the
natural number coordinate of the corresponding point.

For two real numbers $a$ and $b$ the probability
$P_{a\cup b}dr$ of finding $s$ in the union of the
neighborhoods $(dr)_a\cup (dr)_b$
\begin{equation}
P_{a\cup b}\,dr\ne [P(a)+P(b)]\,dr
\end{equation}
because $s$ corresponds to both (all) points at the
same time (the elemental events are not mutually exclusive).
In other words, the probability is inevitably non-additive.
In order to overcome this obstacle,
it is most natural to introduce a function $\psi(r)$
such that $P(r)={\cal P}[\psi(r)]$ and
$\psi_{a\cup b}=\psi(a)+\psi(b)$.
The idea is to compute the non-additive probability
from some additive object by a simple rule. It is quite
clear that this rule should be non-linear. Indeed,
\begin{equation}
P_{a\cup b}={\cal P}(\psi_{a\cup b})={\cal P}[\psi(a)+\psi(b)]\ne
{\cal P}[\psi(a)]+{\cal P}[\psi(b)],
\end{equation}
i.e., the dependence ${\cal P}[\psi (r)]$ is non-linear.
We may choose the dependence arbitrarily but the simplest
option is always preferable.
The simplest non-linear dependence is the square dependence:
\begin{equation}
{\cal P}[\psi(r)]=|\psi(r)|^2.
\end{equation} \label{born}
Of course, we must be ready to correct our choice.
Therefore, running a little ahead, we providently use
modulus brackets: the function $\psi(r)$ will be complex-valued in order
to ensure invariance of the probability under shift in $N$. The square
dependence will also lead to physically clear concept of wave
which is the important additional reason for the choice.

We shall not discuss uniqueness of the chosen options.
The aim of this paper is to show that quantum mechanics is,
at least, one of the simplest and most natural descriptions
of the set of intermediate cardinality.

The probability $P(r)$ is not
probability density because of its non-additivity.
This fact is very important.
Actually, the concept of probability should be modified,
since the additivity law is one of the axioms of the
conventional probability theory (the sample space should consist
of the mutually exclusive elemental events).
But we shall not alter the concept of probability because it is not
altered in quantum mechanics. 
This means that we shall regard $P(r)=|\psi(r)|^2$ as probability
density, i.e., we accept an analogue of Born postulate. 

The function $\psi$, necessarily, depends on $n$: $\psi(r)\to\psi(n,r)$.
Since $n$ is accurate up to a constant (shift)
and the function $\psi$ is defined up to the 
factor $e^{i\mbox{const}}$, we have
\begin{equation}
\psi (n+\mbox{const},r)=e^{i\mbox{const}}\psi (n,r).
\end{equation}
Hence, the function $\psi$ is of the following form:
\begin{equation}
\psi (r,n)=A(r)e^{2\pi in}\label{debr}.
\end{equation}

Thus the point of the intermediate set corresponds to
the function Eq.(\ref{debr}) in continuum. We can specify
the point by the function $\psi(n,r)$ before the mapping
and by the random real number and the natural
number when the mapping has performed. In other words,
the function $\psi(n,r)$ may be regarded as the image
of $s$ in $R$ between mappings.

Consider probability $P(b,a)$ of finding the point $s$ at $b$
after finding it at $a$.
Let us use a continuous parameter $t$ for correlation
between continuous and countable coordinates of the point
$s$ (simultaneity) and in order to distinguish between
the different mappings (events ordering):
\begin{equation}
r(t_a),n(t_a)\to\psi(t)\to r(t_b),n(t_b),
\end{equation}\label{0-t}
where $t_a<t<t_b$ and $\psi(t)=\psi[n(t),r(t)]$.
For simplicity, we shall identify the parameter with
time without further discussion. Note that we cannot use the
direct dependence $n=n(r)$. Since $r=r(n)$ is a random number,
the inverse function is meaningless.

Assume that for each $t\in (t_a,t_b)$ there exists the image of
the point in continuum $R$. 

Partition interval $(t_a,t_b)$ into $k$ equal parts
$\varepsilon$:
\begin{eqnarray}
k\varepsilon =t_b-t_a,\nonumber\\
\varepsilon =t_i-t_{i-1},\nonumber\\
t_a=t_0,\,t_b=t_k,\\
a=r(t_a)=r_0,\, b=r(t_k)=r_k.\nonumber
\end{eqnarray}\label{partition}
The conditional probability of of finding the point $s$ at
$r(t_i)$ after $r(t_{i-1})$ is given by
\begin{equation}\label{cond}
P(r_{i-1},r_i)=\frac{P(r_i)}{P(r_{i-1})},
\end{equation}
i.e.,
\begin{equation}
P(r_{i-1},r_i)=\left|\frac{A_i}{A_{i-1}}e^{2\pi i\Delta n_i}\right|^2,
\end{equation}
where $\Delta n_i=|n(t_i)-n(t_{i-1})|$.

The probability of the sequence of the transitions 

\begin{equation}
r_0,\ldots ,r_i,\ldots r_k
\end{equation}\label{sequence}
is given by
\begin{equation}
P(r_0,\ldots ,r_i,\ldots r_k)=\prod_{i=1}^k P(r_{i-1},r_i)=
\left|\frac{A_k}{A_0}\exp 2\pi i\sum_{i=1}^k\Delta n_i\right|^2.
\end{equation}
Then we get probability of the corresponding continuous sequence
of the transitions $r(t)$:
\begin{equation}\label{pathprob}
P[r(t)]=\lim_{\varepsilon\to 0}P(r_0,\ldots ,r_i,\ldots r_k)=\left|\frac{A_k}{A_0}e^{2\pi im}\right|^2,
\end{equation}
where
\begin{equation}
m=\lim_{\varepsilon\to 0}\sum_{i=1}^k\Delta n_i.
\end{equation}

Since at any time $t_a<t<t_b$ the point $s$ corresponds to all points
of $R$, it also corresponds to all continuous random sequences of
mappings $r(t)$ simultaneously, i.e., probability $P[r(t)]$ of finding
the point at any time ${t_a\leq t\leq t_b}$ on $r(t)$ is non-additive too.
Therefore, we introduce an additive functional $\phi[r(t)]$.
In the same way as above, we get
\begin{equation}
P[r(t)]=|\phi[r(t)]|^2.
\end{equation}\label{}

Taking into account Eq.(\ref{pathprob}), we can put
\begin{equation}\label{phi}
\phi[r(t)]=\frac{A_N}{A_0}e^{2\pi im}=\mbox{const}\,e^{2\pi im}.
\end{equation}
Thus we have
\begin{equation}\label{pathsum}
P(b,a) = |\!\!\sum_{all\,r(t)}\!\!\mbox{const}\,e^{2\pi im}|^2,
\end{equation}
i.e., the probability $P(a,b)$ of finding the point $s$ at $b$
after finding it at $a$ satisfies the conditions of Feynman's approach
(section 2-2 of \cite{Feynman}) for $S/\hbar=2\pi m$. 
Therefore,
\begin{equation}
P(b,a)=|K(b,a)|^2,
\end{equation}
where $K(a,b)$ is the path integral (2-25) of \cite{Feynman}:
\begin{equation}\label{pathint}
K(b,a)=\int_{a}^{b}\!e^{2\pi im}D r(t).
\end{equation}
Since Feynman does not essentially use in Chap.2 that $S/\hbar$ is just
action, the identification of $2\pi m$ and $S/\hbar$ may be postponed.

In section 2-3 of \cite{Feynman} Feynman explains how
the principle of least action follows from the dependence
\begin{equation}\label{sum}
P(b,a)= |\!\sum_{all\,r(t)}\!\!\mbox{const}\,e^{(i/\hbar)S[r(t)]}|^2:
\end{equation}
``The classical approximation corresponds to the case that ... the phase
of the contribution $S/\hbar$ is some very, very large angle. The real (or
imaginary) part of $\phi$ is the cosine (or sine) of this angle. ... small
changes in path will, generally, make enormous changes in phase, and our
cosine or sine will oscilate exceedingly rapidly between plus and minus
values. The total contribution will then add to zero; ... But for the
special path $\bar x$, for which $S$ is an extremum, a small change in path
produces, in the first order at least, no change in $S$. Therefore, only
for paths in the vicinity of $\bar x$ can we get important contributions,
and in the classical limit we need only consider this particular trajectory
as being of imporance.''
We can apply the same reasoning to Eq.(\ref{pathsum}) and,
for very large $m$, get ``the principle of least $m$''.
This also means that for large $m$ the point $s$ has a definite
stationary path and, consequently, a definite continuous coordinate.
In other words, the corresponding interval of the intermediate
set is sufficiently close to continuum (let the interval be
called macroscopic), i.e., cardinality of the intermediate set depends
on its size. Recall that we can measure the size of an interval of
the set only in the unit sets (some packets of points).

Since large $m$ may be considered as continuous
variable, we have
\begin{equation}\label{lim}
m=\lim\limits_{\varepsilon\to 0}\sum_{i=1}^k\Delta n_i=\int_{t_a}^{t_b}\!\!dn(t)=\min.
\end{equation}
The function $n(t)$ may be regarded as some function of
$r(t)$: $n(t)=\eta[r(t)]$. It is important that $r(t)$ is not
random in the case of large $m$. Therefore,
\begin{equation}\label{f}
\int_{t_a}^{t_b}\!\!dn(t)=\int_{t_a}^{t_b}\!\frac{d\eta}{dr}\,\dot{r}\,dt=\min,
\end{equation}
where $\frac{d\eta}{dr}\,\dot{r}$ is some function of $r$, $\dot{r}$,
and $t$. This is a formulation of the principle of least action
(note absence of higher time derivatives than $\dot{r}$), i.e.,
large $m$ can be identified with action.

Since the value of action depends on units of measurement, we need
a parameter $h$ depending on units only such that

\begin{equation}
hm=\int_{t_a}^{t_b}\!\! L(r,\dot{r},t)\,dt=S.
\end{equation}

Finally, we may substitute $S/\hbar$ for $2\pi m$ in Eq.(\ref{pathint})
and consider Feynman's formulation of quantum mechanics as a natural
description of the set of intermediate cardinality.

\medskip
Thus the intermediate set is a set of variable infinite cardinality. Taking
into account that any infinite set should be equivalent to its proper
subset, we get that the set should have constant cardinality ranges, i.e.,
intermediate cardinality changes stepwise. Addition of only large enough
``portion'' of points changes cardinality of an intermediate subset to the
next level. It is reasonable to identify these portions with the unit sets.

If we reduce size of the large intermediate interval, its length
becomes unstable and then collapses, i.e., we have three basic kinds of
the interval:

Macroscopic interval. This interval is large enough to be regarded as
continuos. It has stable non-zero length.

Microscopic interval. This interval may not be regarded as
continuos. It has no length, i.e., its continuos image is
exactly a point. 

Submicroscopic interval. It is an intermediate
kind of the interval with unstable random length. Its length is
either zero or non-zero random real number depending on mapping.
This property is just wave-particle duality: the submicroscopic
interval looks either like a point or like a continuous interval.
Due to the factor $e^{2\pi im}$ in Eq.(\ref{phi}), this instability
shows periodic character and may be described by means of the concept
of wave. Oscillatory instability makes the difference between the
``classical'' inexact length and quantum random length (and, in
combination with non-additivity, between classical and quantum
probabilities).
Submicroscopic intervals make the region of quantum mechanics.

It is not correct to consider the subset of intermediate
cardinality as a set of random numbers or pairs of
real and natural numbers. We cannot (and we do not need to) exhaust the
entire intermediate set. It is sufficient to find one interval with
unstable length in order to state that we have approximate continuum
(the intermediate set). Since in the set of the real numbers properties
of an interval do not depend on the size of the interval, in exact
continuum, unlike approximate one, the members of the intermediate subset
are somewhat artificial. The intermediate subset in exact continuum
should be regarded as a system of unstable formations in the natural
structure of the real numbers. This system is described by quantum
mechanics.   

Note that we can substitute action for $m$ only for sufficiently
high time rate of change of the countable coordinate $n$ because,
if $\Delta n_i=|n(t_{i})-n(t_{i-1})|$ in Eq.(\ref{lim}) is not
sufficiently large to be considered as an (even infinitesimal)
interval of continuum, action reduces to zero. In other words, the
change in size of the intermediate set, from $t_a$ to $t_b$, should
not be microscopic (exact zero, from macroscopic point of view).
Zero-action may be understood as vanishing of mass of the point.
Recall that mass is a factor which appear in Lagrangian of a
free point as a peculiar property of the point under consideration,
i.e., formally, mass may be regarded as a consequence of
the principle of least action \cite{mech} and, consequently, of
the sufficiently high time rate of change of the countable
coordinate (cardinality). Figuratively, mass is something like air
drag which is substantial only for sufficiently fast bodies.
Note some analogy with the Higgs mechanism: Higgs field also plays
the role of a selective drag factor.

Consider the special case of constant time rate of change $\nu$ of
the countable coordinate $n$. We have $m=\nu(t_b-t_a)$. Then
``the principle of least $m$'' reduces to
``the principle of least $t_b-t_a$''.
If $\nu$ is not sufficiently large (massless point), this is the
simplest form of Fermat's least time principle for light. The more
general form of Fermat's principle follows from Eq.(\ref{lim}): since
\begin{equation}
\int_{t_a}^{t_b}\!\!dn(t)=\nu\!\int_{t_a}^{t_b}\!\!dt=\min,
\end{equation}
we obviously get
\begin{equation}\label{fermat}
\int_{t_a}^{t_b}\!\!\frac{dr}{v(t)}=\min,
\end{equation}
where $v(t)=dr/dt$.
In the case of non-zero action (mass point), the principle
of least action and Fermat's principle ``work''
simultaneously. It is clear that any additional factor
can only increase the pure least time. As a result,
$t_b-t_a$ for a massless point bounds below $t_b-t_a$
for any other point and, therefore, $(b-a)/(t_b-t_a)$
for massless point bounds above average speed between 
the same points $a$ and $b$ for continuous image of any
point of the intermediate set.
This is a step towards special relativity.

Galileo's relativity principle may be considered as a consequence of
replacement, in the macroscopic case, of the absolute natural number $m$
by the integral (action): the integrand (Lagrangian) is defined up to
the total time derivative. This lack of uniqueness results in the
relativity principle. In a literal sense, microscopic absolute space
turns to macroscopic relative space. 

Thus, paradoxically, light consist of the slowest points.
Note that any spacetime interval ${\tau^2=(c\Delta t)^2-(\Delta r)^2}$
along a light beam is exact zero (lightlike or null interval). In this
sense, a photon, in Minkowski spacetime, is at absolute rest: it moves no
spacetime ``distance'' (pseudo-distance). Recall that spacetime interval is
directly related to relativistic action ${S_{rel}\sim\int d\tau}$ and,
consequently, to cardinality of the corresponding path. From this fact,
in particular, it follows that entangled photons are really very close to
each other, in the sense of the countable coordinate, until the measurement
has performed, i.e., we have the proper microscopic interval which has
macroscopic image. The submicroscopic stage is skipped in this case (which
is usual in optical phenomena), therefore, we do not have random real
number. But the direction of the point-like microscopic interval is
independent of the direction of the real line and should be generated
randomly by the measurement.

On the contrary, a point with mass should, permanently, have sufficiently
high time rate of change of its path cardinality. As a consequence, the
continuous image of the point also cannot be fixed. This quite conforms
with the uncertainty principle and seems to be its most important aspect.

\medskip
Since the intermediate set is a set of variable infinite cardinality, there
is no need of an external continuous container-set in order to satisfy the
basic conclusion that the set of intermediate cardinality must be a subset of
continuum. The intermediate set is contained in its own sufficiently large
interval. 

Only sufficiently small intermediate interval manifests explicit
features of intermediate cardinality. In other words, the intermediate set
is a substantially microscopic set. Separation of the intermediate subset
from continuum, in some figurative sense, means ``enlargement'' of the
subset which in turn means increasing of its cardinality and, as a
consequence, loss of ``microscopicity'' of the set, i.e., the
separation transforms the subset ``beyond recognition''.

\medskip
There are two kinds of continuum: exact mathematical (formal)
continuum and approximate real continuum. Formal continuum is
highly homogeneous set: its arbitrarily small interval has the same
cardinality (number of points) as the entire real line: ${|\{dr\}|=|R|}$.
Here ${|\{dr\}|}$ is the number of points of the infinitesimal interval
$dr$. The infinite number $|R|$ of intermediate points between two converging
points of exact continuum never decreases until the points have coincided
and the set of these points becomes empty at once. This leap is analogous to
the wave function collapse (conversion of the continuous image of the
submicroscopic interval).

As a consequence of its super-homogeneity, formal continuum is static:
it does not possess the principle of least action. Note that geometric
figures are motionless and massless. Motion, mass, and physical laws
should be imported into formal continuum from outside. Motion is a
consequence of the structure difference in the intermediate set.

In fact, physics for a long time has the need of a set of variable
infinite cardinality. Physical properties of matter vary with its size
reduction whereas properties of formal continuum remain invariable.
This is very inconvenient and absolutely unrealistic if the points of
continuous space have any physical meaning.

\medskip
It is important to make some general qualitative remarks about the
complete description of the set of intermediate cardinality.

The description of a point in the intermediate set depends on the time
rate of change of its path cardinality $m$.

A sufficiently fast point produces the macroscopic path. It has
definite continuous coordinate on the path.
Since the principle of least action is an intrinsic property of the
set of intermediate cardinality relating to the macroscopic paths
in the set, it may be stated that classical mechanics is a description
of the point on the macroscopic intermediate interval.

Quantum mechanics describes the point on the submicroscopic interval in
terms of the continuous description. This description also has its
intrinsic law: the wave equation.

From macroscopic point of view, there are two kinds
of points: the true points and the composite points. A composite
point (the microscopic interval) consist of an infinite number of
points. It is uniquely determined by the natural number of unit sets.
Cardinality of the proper microscopic interval may be regarded as
some qualitative property of the point (charge).
If the interval is destroyed (decay of the corresponding point),
this property vanishes and turns to the properties (cardinalities)
of the output intervals. (We may formulate the following marginal
thesis: qualitative properties are infinite quantities.)

Thus the description of the proper microscopic intervals reduces to
the description of transmutation of expanded (non-local) but, at the
same time, point-like objects and their properties.

The string theory clearly shows that particle properties are really
properties of some intervals and segments having natural numbers
as their inherent characteristics.

We see that the complete description of the set of intermediate
cardinality falls into a chain of three theories. Each theory
corresponds to a class of approximately equivalent intervals
(scale).

Non-equivalent macroscopic, submicroscopic, and microscopic
intervals of the set of intermediate cardinality may not be regarded
as the same homogeneous axis. As a result the description of a true
point in the one-dimensional intermediate set is three-dimensional
(or four-dimensional, including time).

Each of the three descriptions has its particular main law, directions,
and dimensions. Thus, in addition to teleportation, we have parallel
worlds\footnote{By the way, isn't it clear that classical and quantum
mechanics describe different worlds?}.
This worlds are partially autonomous because they are self-contained
structures consisting of different unmatched (``inadherent'') components
(``assembly units''). The proper microscopic world consist of the true zero-length
points. The submicroscopic world consist of unstable (oscillating) intervals
(oscillators of quantum field theory). The points of the macro-world are,
actually, infinitesimal continuous intervals (the least intervals with
sufficiently stable length): any sum (integral) of exact zero values is
exact zero and therefore zero-length points can not constitute the real
line or its interval. Recall that, in set theory, uncountable sets appear
only due to the power set axiom.

The absolute sizes of the intervals (degree of the intervals size reduction)
determine its generic properties and the total number of the intervals
(cardinality of the set they constitute). This obvious interdependence
establishes the connection between cardinality and structure of an infinite
set (the second thesis). Note that the way of obtaining sets by the fission
of a whole is diametrically opposite to the step-by-step construction which
is the preferable (in fact, the only admissible) way in set theory.

Due to the absence of strict correlation of the worlds,
the interactions between them need special arrangements (interfaces).
Macroscopic measuring apparatus forms interface between the macroscopic
and submicroscopic worlds. Interactions between the descriptions is one
of the most unclear points of the consideration.

At present, all the descriptions are imbedded in the continuous space of
classical mechanics. As a result, the dimensions and the directions
of the submicroscopic and microscopic descriptions are lost.

The total number of space time dimensions of three 3D descriptions
is ten. The same number of dimensions appear in string theories.
But the extra dimensions of the intermediate set are essentially
microscopic and do not require compactification.

The directions of the submicroscopic and microscopic descriptions
are replaced with spin. Reliable separation of the descriptions
needs careful examination but it may be preliminarily stated that
integer spin is the direction of the microscopic description and
half integer spin is the direction of the submicroscopic one.
Since the submicroscopic interval is the (unstable) continuous
interval, its direction is associated with the direction of the
macroscopic continuous interval. Therefore, the submicroscopic
direction is not a vector of full value but only spinor.

The direction of the point-like proper microscopic interval is
independent of the continuous direction.

A microscopic interval has structure induced by its cardinality. 
Existence of the ranges of constant cardinality makes possible equivalence
relations (symmetries) inside these ranges which determines applicability
of symmetry groups. Therefore, cardinality of a set may be characterized by
type of symmetry. Outside a subset of certain cardinality its internal
symmetry should be broken. As a result, the complete description should
split into ``asymmetric'' parts (different theories) subject to the number
of microscopic scales (distinguishable cardinalities). Non-equivalent
subintervals form additional microscopic extra dimensions down to the single
unit set. The similar situation is considered as a disadvantage of string
theory.

Algebraic structure of space (in the physical sense of the word) is the
remainder and, at the same time, a framework of its geometric structure.

The answer to the question ``why fermions and bosons obey
different statistics?'' may be very short: ``because they belongs
to the different descriptions''. The Pauli exclusion principle
is a condition for keeping inside the submicroscopic description
(in other words, this is just a condition of conservation of
submicroscopic cardinality): if two points at a submicroscopic
distance come close enough, in the sense of the countable coordinates,
they form the proper microscopic interval and go over to the proper
microscopic description. In this case, some macroscopic and
submicroscopic properties of the points of the interval may be lost.

Each microscopic scale should have analogous condition of conservation
of its cardinality, i.e., the law of conservation of some qualitative
property (charge). Violation of this law means conversion of initial
cardinality into cardinality of another scale.

\medskip
It is important to stress that the intermediate set is not one more
configuration (auxiliary) space. Intermediate cardinality $|I|$ is an
infinite fundamental constant: the number of all points in the Universe.
This constant is the main experimental result of quantum physics.
It is equally valid for micro-physics, macro-physics, and cosmology.
Note that such a global parameter can be determined only in sufficiently
small region. 

Experiments at particle accelerators, on the supposition of continuous
space, can be compared to the launching of space vehicles in the framework
of Ptolemaic geocentric system. Ptolemaic system was based on the false
space model but, long before harmonic analysis, Ptolemy successfully
approximated real planetary motion by a combination of circular motions
(epicycles). Modern micro-physics is based on combining symmetries. Such a
description may be considered as some kind of generalization of harmonic
analysis. This is a sign of approximate model based on false space concept.
No wonder that the complete visual space picture of microscopic phenomena
seems impossible and unnecessary: visual description of full value
cannot be obtained from a false model anyhow.

Note that there is a reason for using symmetries besides
making an approximate model: existence of subintervals of constant 
cardinality in the absence of geometry and stable continuous metrics.
However, symmetries suitable for approximate model are different from
symmetries of the real spatial structure as well as epicycles differ
from the real planetary orbits.
The real symmetries make possible approximate description.

Determination of signs of approximate model based on wrong elementary
concepts, in contrast to the correct model, is rather subject of model
theory. Obviously, such an approximate model should be fragmented and
incomparably more complicated than the true model.

Since model theory does not have concept of
approximate model, consider one more brick road illustration:
Let the brick road consist of only black-and-white bricks (top
side is completely black, bottom side is completely white or vice
versa). We can easily get the bottom view of the road from its top
view. But if one is sure that the road consist of the completely
black and completely white bricks, obtaining of the bottom view is not
always possible. If the top view of the one-dimensional infinite brick
road consist of alternating black and white sides, we can get
the bottom view by the one-brick shift. The obtaining is obviously
possible by means of certain transformations in case of some other
overall symmetries. But if the road is infinite and the arrangement of
the bricks is absolutely random, the problem is, in general,
undecidable. We also get the complicative requirement: in the case of
monochromatic bricks, the number of black (white) sides in the bottom
view must be equal to the number black (white) sides in the top view.
If this ``supersymmetry '' breaks, we should invent some ``dark bricks'' for
matching of the views. We may also make two different descriptions for
top and bottom views separately (which will appear to be dual because
the arrangement of the black (white) bricks in the description of the top
view coincides with the arrangement of the white (black) bricks in the
description of the bottom view) or find and use accidental local symmetries.

Thus, if one tries to describe the road made of black-and-white bricks
by the ``notions'' of the completely black and completely white bricks,
one will inevitably get fragmented description based on symmetries
vaguely related to the symmetries of the true arrangement of the bricks,
which, in principle, cannot be unified into consistent spatial picture.
In the general case, such a description has infinite complexity.
This is the cost of incorrect choice of the basic notions.

\medskip
As outlined above, the consistent visual space picture of microscopic
phenomena is possible at least in principle, although the complete
description becomes complicated by the microscopic and submicroscopic
``sub-worlds''. Consistent visualization is a very strong argument.
Note that, in classical mechanics, a theoretic scheme always obeys the
visual picture. 

The physical description of nature falls into a collection of different
theories steadily resisting  unification. The complete description of
the intermediate set exhibits the same tendency.
This is a consequence of the inherent structural nonuniformity of the set.
The theory of everything seems to be an unreal concept analogous to the
self-contradictory concept of the set of all sets. It is important to note
that this description (or rather the system of descriptions) follows from
the only fact: space is neither continuous nor discrete.

Thus instead of expected new fundamental principles of GUT or M-theory,
we get a new fundamental object. The different
fundamental theories appear to be the legitimate component parts of
its complete description. One should admit that this is more
correct way to unification: we get real properties instead of
formal schemes. These properties will not be abandoned when the
investigation will go deeper into the structure of matter.
The formal introduction of new more and more complicated properties of
the same old objects (waves, particles, strings and membranes,
dimensions, fields, etc) is a deadlock.

In fact, the main logical error of the founders of quantum mechanics
was not indeterminism but the loss of the object under consideration.


\begin{thebibliography}{99}
\bibitem{Cohen}
Cohen P., Set theory and the continuum hypothesis,
New York: W. A. Benjamin, 1966.

\bibitem{Feynman}
Feynman R. P., Hibbs A. R., Quantum mechanics and Path
Integrals, McGraw-Hill Book Company, New York, 1965.

\bibitem{mech}
Landau L. D., Lifshitz E. M., Mechanics,
Oxford; New York: Pergamon Press, 1976.


\end{thebibliography}
\end{document}